\documentclass[prl,aps,showpacs,twocolumn,floatfix,superscriptaddress]{revtex4}
\usepackage{graphicx}
\usepackage{epsfig}

\newcommand{\beq}{\begin{equation}}
\newcommand{\eeq}{\end{equation}}
\newcommand{\beqs}{\begin{eqnarray}}
\newcommand{\eeqs}{\end{eqnarray}}

\newcommand{\Tr}{{\rm Tr}}

\def\hbar{\hspace{0pt}\raisebox{1pt}{$-$} \hspace{-7pt} h}

\begin{document}
\title{$SU(3)$ Family Gauge Symmetry and the Axion}

\author{Thomas Appelquist \thanks{email: thomas.appelquist@yale.edu}}
\author{Yang Bai \thanks{email: yang.bai@yale.edu}}
\affiliation{Department of Physics, Sloane Laboratory, Yale University, New
Haven, CT 06520}
\author{Maurizio Piai \thanks{email:piai@u.washington.edu }}
\affiliation{Department of Physics, University of Washington, Seattle, WA
98195}

\date{\today}
%\vspace{6mm}

\begin{abstract}

We analyze the structure of a recently proposed effective field theory (EFT)
for the generation of quark and lepton mass ratios and mixing angles, based
on the spontaneous breaking of an $SU(3)$ family gauge symmetry at a high
scale $F$. We classify the Yukawa operators necessary to seed the masses,
making use of the continuous global symmetries that they preserve. One
global $U(1)$, in addition to baryon number and electroweak hypercharge,
remains unbroken after the inclusion of all operators required by
standard-model-fermion phenomenology. An associated vacuum symmetry insures
the vanishing of the first-family quark and charged-lepton masses in the
absence of the family gauge interaction. If this $U(1)$ symmetry is taken to
be exact in the EFT, broken explicitly by only the QCD-induced anomaly, and
if the breaking scale $F$ is taken to lie in the range $10^{9} - 10^{12}$
GeV, then the associated Nambu-Goldstone boson is a potential QCD axion.

\end{abstract}

\pacs{ 12.15.Ff, 14.60.Pq, 14.80.Mz }

\maketitle

%%%%%%%%%%%%%%%%%%%%%%%%%%%%%%%%%%%%%%%%%%%%%%%%%%%%%%%%%%%%%%%%%%%%%%
%{\em \bf{Introduction}---}%
%%%%%%%%%%%%%%%%%%%%%%%%%%%%%%%%%%%%%%%%%%%%%%%%%%%%%%%%%%%%%%%%%%%%%%
\section{Introduction}

In a recent set of papers~\cite{APY2}~\cite{APY3}, we developed an
effective-field-theory (EFT) framework for the computation of quark and
lepton masses and mixing angles based on an $SU(3)_F$ family gauge symmetry.
The largest elements of the quark and charged-lepton mass matrices are
seeded phenomenologically through a set of Yukawa operators, bilinear in the
quark and lepton fields and including the Higgs doublet. They also include
standard-model (SM)-singlet scalars transforming as sextets under the
$SU(3)_F$ family group. The family symmetry is broken spontaneously at a
high scale $F$ by vacuum expectation values (VEV's) of these scalars. The
$SU(3)_F$ family symmetry is realized nonlinearly among the SM-singlet
scalars, so that only Nambu-Goldstone (NGB) and pseudo-Nambu-Goldstone
(PNGB) degrees of freedom remain in the EFT.

The small charged-lepton mass ratios, and the small up-type and down-type
quark mass ratios and Cabibbo-Kobayashi-Maskawa (CKM) mixing angles are then
computed perturbatively in the family gauge coupling. Small, hierarchical
neutrino masses, and large leptonic mixing angles are naturally accommodated
at zeroth order in the family gauge coupling, although the specific values
of the mixing angles are not predicted ~\cite{APY3}. Imposing the
constraints from the measured solar and atmospheric mass differences and
mixing angles restricts the parameters describing the vacuum symmetry
structure and can relate some of the otherwise free parameters in the quark
and charged-lepton mass-matrix estimates. One small leptonic mixing angle
then emerges, and is predicted to lie within the reach of planned
experiments.

To classify the Yukawa operators of the EFT, we found it helpful
in Refs.~\cite{APY2}~\cite{APY3} to make use of a discrete, $Z_3$,
symmetry. Here we dispense with the $Z_3$ and show that a complete
classification scheme is provided through the set of global $U(1)$
symmetries associated with each of the complex fields of the
model. One combination is rendered anomalous by the $SU(3)_F$
family gauge interaction. We show that the dominant Yukawa
operators required to describe (with the family gauge
interactions) most features of the quark and charged-lepton mass
matrices then preserve two $U(1)$ symmetries in addition to those
associated with baryon number and electroweak hypercharge.

In order to fit precisely the quark and charged-lepton mass matrices, and to
generate the neutrino mass matrix, it is necessary to include some
additional, smaller operators that explicitly break these two $U(1)$
symmetries to one. This final symmetry, $U(1)_a$, is broken spontaneously at
the scale $F$. If it is taken to be exact in the EFT, broken explicitly by
only QCD anomalies, it could play the role of a Peccei-Quinn symmetry to
address the strong CP problem~\cite{PQ}.

We first discuss the model and the Yukawa operators necessary to
seed the quark and lepton mass matrices. We then describe the
approximate global symmetries of the EFT, broken explicitly by the
family gauge interactions and the Yukawa operators. We discuss the
vacuum structure of the EFT, enumerating the NGB's and PNGB's, and
then classify the fermion mass matrices that emerge from the
Yukawa operators. We conclude with a discussion of the $U(1)_a$
global symmetry of the EFT, broken explicitly by QCD anomalies,
and leading to a potential axion~\cite{axion}.

\section{The Model}

The model of Refs.~\cite{APY2}~\cite{APY3} consists of the three
families of SM fermions, together with two additional fermions,
$\chi$ and $\chi^c$, also coming in three families, required to
explain the up-type quark mass ratios. Each of the (left-handed,
chiral) fermion fields, $q, u^c, d^c, \chi, \chi^c, l, e^c$,
transforms as a $\bf{3}$ under a family $SU(3)_1$ symmetry. Two
complex, symmetric-tensor fields $S$ and $\Sigma$ ($\bf{\bar
6}$'s) are employed to seed the spontaneous breaking of the
$SU(3)_1$. These fields constitute the ``visible" sector of the
model. With electroweak symmetry breaking described by a single
Higgs-doublet field, some additional mechanism is required to
stabilize the Higgs mass. This problem was not addressed in
Refs.~\cite{APY2}~\cite{APY3}, and will not be addressed here.

In order to compute the small quark mass ratios $m_d / m_b$, $m_s
/ m_b$, $m_u /m_t$, $m_c / m_t$, and the CKM mixing angles
radiatively in the family gauge interaction, these quantities must
vanish in its absence. To this end, a ``hidden sector" is
introduced transforming according to its own $SU(3)_2$. The
$SU(3)_F$ family gauge interaction then arises from gauging the
diagonal subgroup of $SU(3)_1 \times SU(3)_2$.

The family breaking scale $F$ is taken to be large enough to suppress
flavor-changing neutral currents, and the family gauge coupling $g$ is weak
enough so that the gauge-boson masses, of order $gF$, are small compared to
the cut-off $M_F = 4 \pi F$ of the EFT. Their effects can therefore be
computed perturbatively within the EFT. The vanishing of the above mass
ratios and the CKM angles in the absence of the family gauge interaction
follows from the symmetries and vacuum structure in the visible sector.
These symmetries are then broken in the hidden sector, with the breaking
communicated to the visible sector through the gauge interactions, leading
to nonzero, calculable values for the mass ratios and CKM angles.

The matter-field content of the EFT is summarized in Table~\ref{tab:fields}.
The hidden sector is described by a single complex, symmetric tensor field
$H$, transforming as a $\bf{\bar 6}$ under $SU(3)_2$. Note that no
SM-singlet neutrinos are included in the EFT. If they exist, they are taken
to have masses above the cutoff $M_F$, and have been integrated out. The EFT
includes the fermion fields, the NGB and PNGB components of $S$, $\Sigma$
and $H$, the family gauge fields, and SM gauge fields.

The $SU(3)_F$ family gauge interaction is, so far, anomalous,
requiring the existence of additional heavy fermions to remove the
anomalies. An example is a set of three SM-singlet fermions, each
transforming as a $\bf{\bar 6}$ under $SU(3)_2$.  With these
``hidden-sector" fermions coupled to $H$, they all become massive
when $H$ develops its symmetry-breaking VEV of order $F$. If their
Yukawa couplings are strong ($O(4\pi)$), then the masses will be
$O(M_F = 4 \pi F)$, and they will not be part of the EFT. When
integrated out, they generate an appropriate Wess-Zumino-Witten
(WZW) term at energies below $M_F$ ~\cite{WZW}. It must be
included in the EFT, but it does not affect the mass estimates of
Refs.~\cite{APY2}~\cite{APY3} to leading order.

              \begin{table}
\begin{tabular}{||c|c|c|c|c|c||}
               \hline
               % after \\: \hline or \cline{col1-col2} \cline{col3-col4} ...
                & $SU(3)_1$ & $SU(3)_2$  & $SU(3)_c$ & $SU(2)_L$ & $U(1)_Y$
                \\ \hline \hline
              $q$  & 3 & 1 &   3 & 2 & $\frac{1}{6}$ \\ \hline
              $u^c$  & 3  &1    & $\bar{3}$ & 1 & $-\frac{2}{3}$ \\ \hline
              $d^c$  &3  & 1   & $\bar{3}$ & 1 & $\frac{1}{3}$ \\ \hline
              $\chi$  &  3 & 1 & 3 & 1 & $\frac{2}{3}$ \\ \hline
              $\chi^c$  & 3 & 1  & $\bar{3}$ & 1 & $-\frac{2}{3}$ \\ \hline
              $\ell$ & 3 & 1 & 1  & 2 & $-\frac{1}{2}$\\ \hline
              $e^c$ & 3 & 1   & 1 & 1 & 1 \\ \hline \hline
              $h$  & 1 &  1 &  1 & 2 & $-\frac{1}{2}$ \\  \hline
              $S$    & $\bar{6}$ & 1   & 1 & 1 & 0 \\  \hline
              $\Sigma$  & $\bar{6}$ & 1   & 1 &  1& 0 \\  \hline  \hline
              $H$  & 1 & $\bar{6}$ & 1   & 1 & 0 \\ \hline
               \hline
\end{tabular}
\caption{Field content and symmetries of the model.
      All fermions are LH chiral fields. The
symbols $S$, $\Sigma$, and $H$ denote SM-singlet scalar fields.}
\label{tab:fields}
\end{table}

\section{Yukawa Operators of the EFT, and $U(1)$ symmetries}

\subsection{Dominant Yukawa Operators}

In the absence of the Yukawa operators, there exists a $U(1)$
global symmetry for each of the $11$ complex fields of
Table~\ref{tab:fields}. A minimal set of  Yukawa operators
required to seed most features of the quark- and charged-lepton
mass matrices is given by \beqs \label{LagY} -{\cal L}_Y &=&
y_d\frac{qhSd^c}{F}+y_1\frac{q\tilde{h}S\chi^c}{F}+y_2\chi
Su^c+y_3\chi\Sigma \chi^c\nonumber\\
&&+ y_e\frac{\ell hSe^c}{F}+{\rm h.c.}. \eeqs The dimensionless
coupling constants, fit to experiment, range in size from
$O(10^{-3})$ to $O(1)$, with electroweak symmetry breaking arising
from the Higgs VEV $v \simeq 250$ GeV. The VEV's of $S$ and
$\Sigma$ are of order $F$. (The $H$ field, so far not directly
coupled to the visible sector, also develops a VEV of order $F$.)
The first and last terms seed the largest elements of the
down-type and charged-lepton mass matrices. The other three terms
are required to set up a (``see-saw") mass-generating mechanism in
the up-type sector~\cite{APY2}. All these operators are
dimension-$3$ or $4$ in the fields with SM quantum numbers.

The phenomenological consequences of these operators were analyzed in
Refs.~\cite{APY2}~\cite{APY3}. There are many other Yukawa operators allowed
by the SM symmetries and the $SU(3)_F$ gauge symmetry, especially since the
$SU(3)_F$ symmetry is realized nonlinearly in the scalar ($S$, $\Sigma$, and
$H$) sectors. In order to justify using only these operators we will make
use of the $U(1)$ symmetries that are naturally part of the model.

The $5$ operators of ${\cal L}_Y$ break $5$ of the $10$ $U(1)$
symmetries associated with the visible-sector fields of
Table~\ref{tab:fields}. In addition, one combination, which can be
taken to be lepton number, $U(1)_{\ell}$, is rendered anomalous by
the $SU(3)_F$ family gauge interaction. Of the remaining $4$
$U(1)$'s, $2$ are $U(1)_B$ corresponding to baryon number and
$U(1)_Y$ corresponding electroweak hypercharge. The final $2$ are
denoted $U(1)_a$ and $U(1)_b$. We exhibit in Table~\ref{tab:U(1)}
one possible choice for the charge assignments of each of the
complex fields under $U(1)_a \times U(1)_b$. The reason for the
charge assignments of $H$ will be made clear shortly.

We will show using the vacuum structure of the EFT that the
operators of ${\cal L}_Y$ provide the required dominant seeding of
the quark and charged-lepton mass matrices, that is, that other
Yukawa operators respecting the $U(1)_a \times U(1)_b$ symmetry
provide no new mass-matrix structure.

              \begin{table}
\begin{tabular}{||c|c|c|c|c|c|c|c|c|c|c|c||}
               \hline
               % after \\: \hline or \cline{col1-col2} \cline{col3-col4} ...
      &$q$ &$u^c$ &$d^c$ &$\chi$ &$\chi^c$ &$\ell$ &$e^c$ &$h$ &$S$ &$\Sigma$
&$H$
                \\ \hline \hline
      $U(1)_a$ & 0 & 0 & 2 & 1 & 0 & -11 & 13 & -1& -1& -1& 20       \\ \hline
      $U(1)_b$ & 1 & 0 & 0 & 0 & -2 & -1 & 2 & -1& 0 & 2& 0        \\ \hline
               \hline
\end{tabular}
\caption{Two linearly independent $U(1)$'s, in addition to $U(1)_B$ and
$U(1)_Y$, left unbroken by the operators of ${\cal L}_Y$ and by anomalies
generated by $SU(3)_F$ family gauge interactions. The small operators of
${\cal L}_Y^{\prime}$ break $U(1)_b$ leaving $U(1)_a$ unbroken.}
\label{tab:U(1)}
\end{table}

\subsection{Smaller, Symmetry-Breaking Yukawa Operators}

The $5$ operators of ${\cal L}_Y$ allow us to fit the quark mass
ratios and CKM mixing angles, except for the smallest CKM angle,
$\theta_{13}^q$. Also, there is nothing in the model so far to
generate charged-lepton mass ratios that differ from the down-type
quark mass ratios. Finally, there is no mechanism so far to
provide the very small neutrino masses and leptonic mixing angles.

Each of these problems can be addressed by including a set of
``smaller" operators that explicitly break one or more of the
symmetries preserved so far. A minimal set, employed in
Ref.~\cite{APY3}, is given by  \beqs \label{BreakingLag-tree}
-{\cal L}_Y^{\prime} &=& y_u^{\prime}\frac{q\tilde{h}\Sigma
u^c}{F} + y^{\prime}_e\frac{\ell h\Sigma
e^c}{F}+\frac{y_{\nu}^{\prime}}{2}\frac{\ell\tilde{h}H\tilde{h}\ell}{F^2}+{\rm
h.c.}.
      \eeqs
The first two operators each break $U(1)_b$ but preserve $U(1)_a$. The
phenomenological use of these operators requires that $y_u^{\prime}$ and
$y^{\prime}_e$ be of order $10^{-4}$.

The third operator, dimension-$5$ in the SM fields, couples the hidden and
visible sectors directly. With the charge assignment for $H$ under $U(1)_a$,
shown in Table~\ref{tab:U(1)}, this operator preserves this symmetry. It
breaks $U(1)_H$ to a combination of $U(1)_H$ and $U(1)_{\ell}$, which is
anomalous due to the $SU(3)_F$ family gauge interactions. It also breaks
$SU(3)_1 \times SU(3)_2$ to the diagonal subgroup, as does the family gauge
interaction. With $v \simeq 250$ Gev, we have $y_{\nu}^{\prime} \simeq F/
(10^{15}$~GeV) if the operator is to give the correct order of magnitude for
the neutrino masses. Thus $y_{\nu}^{\prime}$ is of order $y_u^{\prime}$ and
$y^{\prime}_e$ or smaller, providing $F$ is of order $10^{11}$~GeV or
smaller. (The charge assignment of $H$ under $U(1)_b$ is chosen so that the
third operator preserves this symmetry, even though it is already broken by
the the first two operators.)

With $H$ charged under $U(1)_a$, the additional, heavy,
hidden-sector fermions required to remove $SU(3)_F$ gauge
anomalies may also carry $U(1)_a$ charge. An example is the set of
$3$ heavy $\bf{\bar 6}$'s coupled to $H$, discussed above. In
order that the global $U(1)_a$ not be anomalous due to the
$SU(3)_F$ gauge interaction, the $U(1)_a$ charge assignments of
all the fermions will then have to be adjusted relative to the
values in Table~\ref{tab:U(1)}, but nothing in the present paper
depends on these specific values.

The $U(1)_a$ symmetry, unbroken by the operators of ${\cal L}_Y$
and ${\cal L}_Y^{\prime}$ or by $SU(3)_F$-generated anomalies, is
spontaneously broken at the scale $F$ and is rendered anomalous by
QCD interactions. If it is respected by all the operators of the
EFT, it is a candidate for a Peccei-Quinn symmetry. We return to
this topic after discussing the vacuum structure of the EFT and
its consequences for the fermion mass matrices.

\section{Vacuum Structure}

      In Refs.~\cite{APY2}~\cite{APY3}, we assumed that the global symmetries
$SU(3)_1 \times SU(3)_2$ are broken spontaneously at the scale $F$ by VEV's
of the scalar fields $S$, $\Sigma$ and $H$. The VEV's were taken to be \beqs
            \label{tree-vev}
\langle S\rangle & = & F\left(
                   \begin{array}{ccc}
                    0 & 0 & 0 \\
                     0 & 0 & 0 \\
                     0 & 0 & s \\
                   \end{array}
                 \right)\,\nonumber\\
\langle\Sigma\rangle & = & F\left(
                   \begin{array}{ccc}
                     0 & 0 & 0 \\
                     0 & \sigma & 0 \\
                     0 & 0 & 0 \\
                   \end{array}
                 \right)\,\nonumber\\
\langle H\rangle & = & F\left(
                   \begin{array}{ccc}
                   b_1^2 & b_2 & b_3 \\
                    b_2 & a_1 & a_2 \\
                    b_3 & a_2 & a_3 \\
                   \end{array}
                 \right)\,,
                 \eeqs
where $|s|$, $|\sigma|$ and the $|a_i|$ are $ O(1)$, while the $|b_i|$ are
of order the Cabibbo angle $\theta_{12}^q$. This pattern, which was at the
core of the phenomenology of Refs.~\cite{APY2}~\cite{APY3}, is adopted here.
We next discuss the broken symmetries associated with this VEV pattern, and
the associated NGB's and PNGB's.

We first neglect the family gauge coupling and the small operators
of ${\cal L}_Y^{\prime}$. The visible-sector scalars $S$ and
$\Sigma$ are taken to couple strongly in the underlying theory,
transforming according to a single $SU(3)_1$ symmetry, together
with $U(1)_a \times U(1)_b$. The underlying dynamics is assumed to
trigger the spontaneous breaking of this symmetry in the above
pattern, with $\langle S\rangle$ and $\langle \Sigma \rangle$
together leaving two unbroken $U(1)$ symmetries of the vacuum and
producing $8$ NGB's. The vacuum-symmetry generators are linear
combinations of those of $U(1)_a$, $U(1)_b$, and the two diagonal
generators of $SU(3)_1$. The hidden-sector VEV, $\langle
H\rangle$, produces $9$ NGB's. There are therefore a total of $17$
NGB's, with the $9$ arising from the hidden sector decoupled
so-far from the visible sector.

\begin{figure*}[ht!]

\begin{center}
\includegraphics[width=0.9\linewidth]{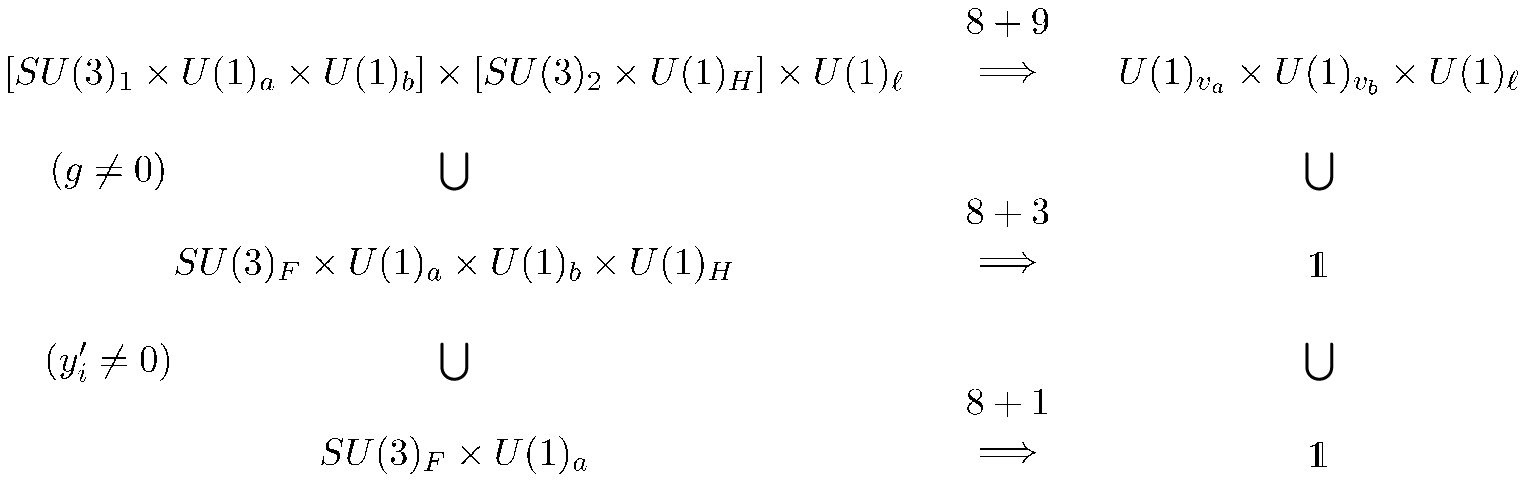} \caption{The symmetry breaking pattern of the model. Horizontal arrows
represent the spontaneous symmetry breaking of Eq.~\ref{tree-vev}. The
numbers over each arrow count the number of NGB's generated. The first row
corresponds to the limit in which the gauge couplings and the small symmetry
breaking operators of ${\cal L}_Y^{\prime}$ are set to zero, in which case
$17$ NGB's are produced. The vacuum symmetries $U(1)_{v_a}$ and $U(1)_{v_b}$
are shown in Eq.~\ref{vac-symm}. In the second row, the $SU(3)_F$ gauge
coupling is turned on, and as a result only the diagonal combination of
$SU(3)_1\times SU(3)_2$ is preserved in the EFT. At the same time, $SU(3)_F$
anomalies explicitly break lepton number $U(1)_{\ell}$. There are 11 NGB's,
and 6 PNGB's with masses $\sim g^2F/4 \pi$. In the third row, the operators
in ${\cal L}_Y^{\prime}$ are turned on, explicitly breaking $U(1)_b$ (by
$y_u^{\prime}$ and $y_e^{\prime}$) and $U(1)_H$ (by $y_{\nu}^{\prime}$).
This gives mass to two of the 11 NGB's. Of these, 8 disappear through the
Higgs mechanism, and one is the candidate axion. Five additional global
$U(1)$'s are explicitly broken by ${\cal L}_Y$, $U(1)_B$ is unbroken, and
$U(1)_Y$ is broken spontaneously at the much lower electroweak scale $v$.
They are not included here. \label{Fig:plot}}
\end{center}
\end{figure*}

In Fig.~\ref{Fig:plot}, we show the symmetry breaking pattern of the model,
with the first line corresponding to the limit in which the gauge couplings
and the operators of ${\cal L}_Y^{\prime}$ are set to zero. The two unbroken
$U(1)$ symmetries of the visible-sector vacuum are designated $U(1)_{v_a}$
and $U(1)_{v_b}$. In the next section (Eq. \ref{vac-symm}), we exhibit these
symmetries explicitly and use them to study the allowed Yukawa operators.

The underlying physics in the visible and hidden sectors, leading to these
patterns, produces a set of nonlinear constraints in the EFT, reducing the
24 degrees of freedom in $S$ and $\Sigma$ to the $8$ NGB's of the visible
sector, and the $12$ degrees of freedom in $H$ to the $9$ NGB's of the
hidden sector. They are described in the Appendix.

%\begin{widetext}

%\end{widetext}

We next include the family gauge coupling and the small operators
of ${\cal L}_Y^{\prime}$. As described above, the family gauge
interaction makes anomalous one visible-sector $U(1)$, taken to be
$U(1)_{\ell}$. Since the family gauge interaction and the third
operator of ${\cal L}_Y^{\prime}$ explicitly break $SU(3)_1 \times
SU(3)_2 \rightarrow SU(3)_F$, since the third operator of ${\cal
L}_Y^{\prime}$ explicitly breaks $U(1)_H$, and since the first two
break $U(1)_b$, the full symmetry of the EFT, excluding the SM
interactions, is $SU(3)_F \times U(1)_a$, along with $U(1)_B$ and
$U(1)_Y$. With the visible and hidden sectors now coupled by the
family-gauge and ${\cal L}_Y^{\prime}$ interactions, the
spontaneous breaking of the $SU(3)_F \times U(1)_a$ symmetry is
complete, producing $9$ NGB's, of which $8$ are eaten. The
remaining NGB is the candidate axion. This count is described in
the last line of Fig.~\ref{Fig:plot}.

Of the original $17$ NGB's, therefore, $8$ have become PNGB's. We
discuss their masses by first noting that if only the family gauge
interaction is included (second line of Fig.~\ref{Fig:plot}), $2$
of the PNGB's, corresponding to the spontaneous breaking of
$U(1)_b$ and $U(1)_H$, remain massless. The other $6$ develop
masses at one-loop in the family gauge interaction, of order $g^2
F / 4 \pi$.

There remain two PNGB's associated with the explicit breaking of $U(1)_b$
and $U(1)_H$ by the operators of ${\cal L}_Y^{\prime}$. In the case of
$U(1)_H$,  a combination of $U(1)_H$ and $U(1)_{\ell}$ is still preserved by
the operators of  ${\cal L}_Y^{\prime}$ (and ${\cal L}_Y$). However, this
($SU(3)_F$-anomalous) symmetry is not an essential ingredient in the quark
and charged-lepton phenomenology, and may be broken by additional operators
which preserve $U(1)_a$, but are small enough not to disturb significantly
the neutrino phenomenology of Ref.~\cite{APY3}. We take these operators to
be present generically. The explicit breaking of $U(1)_b$ by the first two
operators of ${\cal L}_Y^{\prime}$ must be accompanied by a coupling (the
family gauge coupling) between the visible and hidden sectors in order to
make massive the associated NGB.

Consider now the bilinear part of the effective lagrangian involving the
(P)NGB's associated with the spontaneous breaking of $U(1)_a \times U(1)_b
\times U(1)_H$. After proper diagonalization and normalization of the
kinetic part, and neglecting QCD anomalies, there will exist a rank-two mass
matrix generated at the multi-loop level. Its entries are proportional to
$F^2$ with coefficients determined by the small parameters in ${\cal
L}_Y^{\prime}$ and the family gauge coupling $g$. The $2$ resulting PNGB's
(as well as the $6$ PNGB's with masses of order $g^2 F / 4 \pi$) have
non-diagonal couplings to the SM fermions (they are familons), while the
massless NGB has diagonal couplings. In addition, since they all couple
through the third operator of ${\cal L}_Y^{\prime}$ to the Majorana mass
matrix of the neutrinos; they are Majorons. Clearly, $F$ must be taken large
enough so that the $8$ PNGB's lie beyond current experimental reach.

The weak coupling of the visible and hidden sectors by the family
gauge interaction and the third interaction of ${\cal
L}_Y^{\prime}$ means that $\langle H \rangle$ cannot in general be
diagonalized in the frame in which $\langle S \rangle$ and
$\langle \Sigma \rangle$ are diagonal. Its orientation, described
by $3$ mixing angles, is a dynamical, vacuum alignment question.
The mixing angles enter the neutrino mass matrix directly through
the third operator of ${\cal L}_Y^{\prime}$ and they enter the
quark and charged-lepton mass matrices through the $SU(3)_F$
radiative corrections since the gauge-boson mass matrix depends on
$\langle H \rangle$.

The effective potential determining this orientation is generated from the
weak couplings of the EFT as well as other possible weak interactions
linking the two sectors in the underlying theory. To account for the CKM and
Pontecorvo-Maki-Nakagawa-Sakata (PMNS) mixing angles, all the off-diagonal
entries of $\langle H \rangle$ must be non-zero in the basis in which
$\langle S \rangle$ and $\langle \Sigma \rangle$ are diagonal. Equivalently,
the breaking pattern must not leave any residual $Z_2$ symmetries. (The
local operators corresponding to the masses of the PNGB's described above
are a part of the full effective potential.)

Finally, we note that CP-violating phases are naturally present in the
model. They can emerge from the underlying theory and are then present
directly in the Yukawa couplings of the EFT. It can be seen that they cannot
in general be removed from all allowed operators by phase rotations of the
fields. Phases can also arise spontaneously through the weak effective
potential coupling the visible and hidden sectors. The combination of all
these phases will determine the measured CP-violating phase in the CKM
matrix, and the predicted Dirac and Majorana phases in the leptonic (PMNS)
matrix. If the spontaneously generated phases and those present in the
operators of the EFT are $O(1)$, the same will be true of the measured and
predicted phases.

\section{Yukawa Operators and Fermion Mass Matrices}

In this section we examine the phenomenological effects of all
admissible Yukawa operators, including those not in ${\cal L}_Y$
and ${\cal L}_Y^{\prime}$, making use of the EFT symmetries and
the vacuum symmetries. We show that other allowed operators with
the $U(1)_a\times U(1)_b$ symmetry of ${\cal L}_Y$ (or which break
$U(1)_b$ by small amounts as in ${\cal L}_Y^{\prime}$), give no
qualitatively new contributions to the mass matrices at zeroth
order in the family gauge interaction.

The effect of the additional admissible operators we did not include is
therefore at most a redefinition of some of the couplings that seed the mass
matrices, and  hence, even at loop-order in the family gauge interactions,
leave the phenomenological success of ${\cal L}_Y$ and ${\cal L}_Y^{\prime}$
undisturbed.

\subsection{Dominant Yukawa Operators}

We first discuss Yukawa operators respecting the symmetries of
${\cal L}_Y$: $U(1)_a \times U(1)_b \times SU(3)_1$. The family
gauge interaction is initially neglected, it's effects to be
included perturbatively. The operators of interest are bilinear in
the fermion fields and include up to one power of the
Higgs-doublet field $h$. Any number of $S$ and $\Sigma$ fields may
be included since they are subject to the nonlinear constraints
that freeze out all but NGB and PNGB degrees of freedom.

We begin with operators with $S$'s and $\Sigma$'s sandwiched
between $q{\tilde h}$ and $u^c$, that is, operators potentially
capable of directly giving up-type quark masses when the scalars
develop VEV's. If attention is restricted to operators with only
one power of $S$ or $\Sigma$, as in ${\cal L}_Y$, there is no such
quantity. But it easy to write down operators of this type if more
powers of $S$ and $\Sigma$ are admitted. A simple example is
$q{\tilde h}(S \times S)\times \Sigma^{*}u^c / \ F^3$, where $S
\times S$ represents the $\bf 6$ in the product of the two $\bf
\bar{6}$'s. Clearly this operator vanishes in the vacuum of
Eq.~\ref{tree-vev}, but what about the general class of such
operators?

To answer this question, we note that under the $U(1)_{v_a} \times
U(1)_{v_b}$ vacuum symmetry of the visible sector
(Fig.~\ref{Fig:plot}), the fields transform as
      \beqs \label{vac-symm}
      q &\rightarrow&
diag\{e^{i\theta_a},e^{-\frac{1}{2}i\theta_a+2i\theta_b},e^{-\frac{1}{2}i\theta_a+i\theta_b}\}\,q
\nonumber \\
u^c &\rightarrow&
diag\{e^{i\theta_a-i\theta_b},e^{-\frac{1}{2}i\theta_a+i\theta_b},e^{-\frac{1}{2}i\theta_a}\}\,u^c
\nonumber \\
d^c &\rightarrow&
diag\{e^{3i\theta_a-i\theta_b},e^{\frac{3}{2}i\theta_a+i\theta_b},e^{\frac{3}{2}i\theta_a}\}\,d^c
\nonumber \\
\chi &\rightarrow&
diag\{e^{2i\theta_a-i\theta_b},e^{\frac{1}{2}i\theta_a+i\theta_b},e^{\frac{1}{2}i\theta_a}\}\,\chi
\nonumber \\
\chi^c &\rightarrow&
diag\{e^{i\theta_a-3i\theta_b},e^{-\frac{1}{2}i\theta_a-i\theta_b},e^{-\frac{1}{2}i\theta_a-2i\theta_b}\}\,\chi^c
\nonumber \\
\ell &\rightarrow&
diag\{e^{-10i\theta_a-2i\theta_b},e^{-\frac{23}{2}i\theta_a},e^{-\frac{23}{2}i\theta_a-i\theta_b}\}\,\ell
\nonumber \\
e^c &\rightarrow&
diag\{e^{14i\theta_a+i\theta_b},e^{\frac{25}{2}i\theta_a+3i\theta_b},e^{\frac{25}{2}i\theta_a+2i\theta_b}\}\,e^c
\nonumber \\
h &\rightarrow& e^{-i\theta_a-i\theta_b}\,h
\nonumber \\
S &\rightarrow&
diag\{e^{-\frac{3}{2}i\theta_a+i\theta_b},e^{-i\theta_b},1\}\,S\,diag\{e^{-\frac{3}{2}i\theta_a+i\theta_b},e^{-i\theta_b},1\}
\nonumber \\
\Sigma &\rightarrow&
diag\{e^{-\frac{3}{2}i\theta_a+2i\theta_b},1,e^{i\theta_b}\}\,\Sigma\,diag\{e^{-\frac{3}{2}i\theta_a+2i\theta_b},1,e^{i\theta_b}\}
,\nonumber \\
      \eeqs
where $\theta_a$  and $\theta_b$ are the arbitrary parameters associated
with the symmetries $U(1)_a$ and $U(1)_{b}$.

The most general mass operator involving $q$ and $u^c$, emerging
from the VEV's of $S$ and $\Sigma$, is of the form
$q\tilde{h}\,diag\{y_{u_1},y_{u_2},y_{u_3}\}u^c$. In order that it
be invariant under $U(1)_{v_a} \times U(1)_{v_b}$, we must have
      \beqs \label{uptrans}
diag\{y_{u_1},y_{u_2},y_{u_3}\}&=&diag\{e^{3i\theta_a},e^{4i\theta_b},e^{2i\theta_b}\}
      \nonumber \\
       &&\times diag\{y_{u_1},y_{u_2},y_{u_3}\}.
      \eeqs
The only solution is $y_{u_1}=y_{u_2}=y_{u_3}=0$. Thus there is no
Yukawa operator involving $q$ and $u^c$ giving a non-vanishing
mass matrix.

One can show more generally that the mass matrices generated by
the operators of ${\cal L}_Y$ are the most general fermion mass
matrices allowed by $U(1)_{v_a} \times U(1)_{v_b}$. Consider, for
example, a down-type operator with VEV's of $S$ and $\Sigma$
sandwiched between $qh$ and $d^c$. It must be of the form $q
h\,diag\{y_{d_1},y_{d_2},y_{d_3}\}d^c$. In order that it be
invariant, we must have \beqs
      diag\{y_{d_1},y_{d_2},y_{d_3}\}&=&diag\{e^{3i\theta_a -
2i\theta_b},e^{2i\theta_b},1\}
      \nonumber \\
       &&\times diag\{y_{d_1},y_{d_2},y_{d_3}\},
      \eeqs
Thus the only possible non-vanishing entry is the $33$ element,
which is generated by the operator $y_{d}q h S d^c/F$ of ${\cal
L}_Y$. Other operators may be written down that do the same thing,
for example $q h(S \times \Sigma)\times \Sigma^{*}d^c / \ F^3$
with its own (complex) coefficient. It, too, has only a $33$ entry
in the vacuum of Eq.~\ref{tree-vev}.

A similar argument applies to all Yukawa operators respecting the
symmetry $U(1)_a \times U(1)_b \times SU(3)_1$ of the visible
sector. All operators that have non-vanishing VEV's in the vacuum
of Eq.~\ref{tree-vev}, with it's symmetry (Eq.~\ref{vac-symm}),
give rise to the same mass matrices as those arising from the
operators of ${\cal L}_Y$. For the charged-lepton sector there is
only a $33$ entry. For the up-type sector, the entries lay the
groundwork for the see-saw explanation of the masses.

To summarize, we have included in ${\cal L}_Y$ a minimal set of
Yukawa operators necessary to explain, along with the $SU(3)_F$
gauge interaction, most features of the quark and charged-lepton
mass matrices. The two $U(1)$ vacuum symmetries imply that the
quark and charged-lepton mass matrices generated by the operators
of ${\cal L}_Y$ are completely general. Perturbation theory in the
family gauge interaction then couples the visible and hidden
sectors, communicating the breaking of the two $U(1)$ vacuum
symmetries to the visible sector, and leading to non-vanishing
values for up-type and down-type quark mass ratios, CKM mixing
angles, and charged-lepton mass ratios. They are finite and
calculable within the EFT.

\subsection{Smaller, Symmetry-Violating Yukawa Operators}

To incorporate necessary small corrections to the quark- and
charged-lepton mass matrices, and to generate the entire, small
mass matrix of the neutrinos, the additional small operators of
${\cal L}_Y^{\prime}$, are required. The couplings $y_e^{\prime}$
and $y_u^{\prime}$ are $O(10^{-4})$, and $y_{\nu}^{\prime}$ is no
larger than this if $F$ is no larger than about $10^{11}$~GeV.
These operators together break $U(1)_b$,  $U(1)_H$, and $SU(3)_1
\times SU(3)_2 \rightarrow SU(3)_F$, leaving the global $U(1)_a$
symmetry.

With $U(1)_a$ as the only global $U(1)$ symmetry, many other
Yukawa operators, comparably small compared to those of ${\cal
L}_Y^{\prime}$ and breaking $U(1)_b$, are allowed. Note that there
is no distinction between $S$ and $\Sigma$ at this level since
they have the same $U(1)_a$ charges. The question is whether any
of these operators can give rise in the vacuum of
Eq.~\ref{tree-vev} to fermion mass matrices that disturb the
successful phenomenology based on the operators of ${\cal
L}_Y^{\prime}$.

To see that this does not happen, note that the residual symmetry
of the vacuum (present before gauge interactions link the visible
and hidden sectors), with $U(1)_b$ now explicitly broken, is just
$U(1)_{v_a}$. It can be read off from Eq.~\ref{vac-symm} by
setting $\theta_b = 0$. This single $U(1)$ vacuum symmetry allows
nonzero values for only the $22$ and $33$ entries of both $\langle
S \rangle$ and $\langle \Sigma \rangle$. In the absence of the
family gauge interaction, there can therefore be no masses present
for the first-family quarks and charged leptons. This is an
essential role of the $U(1)_a$ symmetry. Its spontaneous breaking
in the hidden sector and transmittal to the visible sector by the
family gauge interaction then produces the small, first-family
masses.

The contributions arising from ${\cal L}_Y^{\prime}$ to the $22$
and $33$ entries of the quark and charged-lepton mass matrices are
$O(10^{-4})$. They produce small but important corrections in the
quark and lepton phenomenology ~\cite{APY3}.

\section{$U(1)_a$ and the QCD Axion}

We have shown that a minimal EFT, capable of accounting for the
quark and lepton masses, mixing angles, and phases
~\cite{APY2}~\cite{APY3}, naturally includes one global $U(1)$
symmetry, $U(1)_a$ of Table~\ref{tab:U(1)}. The breaking pattern
leaves an associated vacuum symmetry $U(1)_{v_a}$
(Eq.~\ref{vac-symm}) in the visible sector, protecting the
first-family quarks and charged leptons from gaining mass in the
absence of the $SU(3)_F$ family gauge interaction. The breaking of
$U(1)_{v_a}$ in the hidden sector at scale $F$, communicated to
the quarks and leptons by the $SU(3)_F$ gauge interaction, leads
to finite first-family masses, and produces a so-far massless NGB.

Suppose next that the $U(1)_a$ symmetry is classically exact, respected by
all operators of the EFT. Then, since it is anomalous due to QCD
interactions, the NGB is a candidate for a QCD axion~\cite{BK}. The axion
field is a linear combination of NGB fields in $S$, $\Sigma$, and $H$, the
combination that remains massless and survives below the scales where the
family gauge bosons and the PNGB's decouple. (This is also below the scale
where the $\chi$ and $\chi^c$ fields have been integrated out, having
generated the up-type quark masses.) The linear combination is dictated by
ratios of the dimensionless parameters that appear in the VEV's of
Eq.~\ref{tree-vev}.

The axion couples to visible matter through the operators of
${\cal L}_Y$ and ${\cal L}_Y^{\prime}$. With the family gauge
corrections included, these operators lead to the observed masses
of all the quarks and leptons. Thus, in the effective theory at
scales low enough so that only the SM fields and the axion
survive, the axion couples to all the quarks and leptons with
coupling strength given by $m_f/F_a$, where $m_f$ is the fermion
mass and $F_a$ is related to $F$ by ratios the dimensionless
parameters that appear in the VEV's of Eq.~\ref{tree-vev}. Since
they are all expected to be roughly of the same order, $F_a$ is of
the same order as $F$. Since the axion couples to neutrino mass
through the Majorana operator in ${\cal L}_Y^{\prime}$, it is also
a Majoron.

It is not our purpose to discuss the phenomenology of this axion candidate
here, except to observe that with $F$ in the allowed window $10^{9} \lesssim
 F \lesssim 10^{12}$~GeV~\cite{pdg,kim}, corresponding to a mass range $10^{-3}
\gtrsim m_a \gtrsim 10^{-6}$~eV, it evades all axion and Majoron searches to
date.

The $U(1)_a$ symmetry is a natural feature of the EFT operators required to
compute quark and lepton mass matrices, and if taken to be exact it leads to
a viable QCD axion. But the imbedding of this EFT in a larger framework
could in general lead to higher-dimension operators that explicitly break
$U(1)_a$ and give contributions to the axion potential that swamp the QCD
contribution~\cite{dine}.

\section {Summary}

We have explored an effective field theory (EFT) framework proposed recently
for the generation of quark and lepton mass matrices
~\cite{APY2}~\cite{APY3}. An $SU(3)$ family gauge symmetry, broken
spontaneously at a high scale $F$, communicates symmetry breaking from a
hidden sector to the visible-sector standard model fields.

To classify the Yukawa operators that seed the mass matrices, we
have employed the set of global $U(1)$ symmetries that are
naturally part of the EFT. The dominant required operators
preserve two such symmetries, $U(1)_a$ and $U(1)_b$, in addition
to baryon number and electroweak hypercharge. A set of smaller
operators, necessary to generate the neutrino mass matrix and to
provide small corrections to the quark and charged lepton mass
matrices, preserve only $U(1)_a$ along with baryon number and
electroweak hypercharge.

We have described the vacuum structure of the EFT, enumerating the
Nambu-Goldstone bosons (NGB's) and pseudo-Nambu Goldstone bosons (PNGB's),
as determined by the symmetry-breaking interactions that link the visible
and hidden sectors. The PNGB's that gain mass because of the family gauge
coupling and the small symmetry-breaking Yukawa operators, couple
off-diagonally in family space (they are familons), and couple to the
Majorana mass matrix of the neutrinos (they are Majorons).

We have used the vacuum structure together with the symmetries of the EFT to
classify the quark and charged-lepton masses that emerge. CP-violating
phases, which lead to the CKM phase as well as Dirac and Majorana phases in
the leptonic PMNS matrix, arise spontaneously within the EFT, and also enter
the parameters of the EFT directly from the underlying physics.

The $U(1)_a$ symmetry is unbroken by any of the phenomenologically
necessary interactions of the EFT, except for QCD anomalies. The
spontaneous breaking pattern preserves an associated vacuum
symmetry, $U(1)_{v_a}$, in the visible sector, enforcing the
masslessness of the first-family quarks and charged leptons in the
absence of the family gauge interaction. The $U(1)_{v_a}$ symmetry
is broken in the hidden sector at the family-breaking scale $F$,
with the breaking communicated to the standard-model fields by the
family gauge interaction.

If the $U(1)_a$ symmetry is taken to be exact in the EFT except
for QCD anomalies (a Peccei-Quinn symmetry), and if $F$ is taken
to lie in the allowed window $10^{9}<F< 10^{12}$~GeV, then the
associated PNGB is a viable axion, coupling to all the particles
of the standard model. This conclusion relies on the large
hierarchy between $F$ and the electroweak scale $v$. Also, it is
not clear whether the $U(1)_a$ symmetry survives the imbedding of
the EFT in a larger framework.

\section {Appendix - Nonlinear Constraints}

We summarize here the nonlinear constraints that must emerge from
the underlying dynamics in the visible sector and the hidden
sector, corresponding to the VEV pattern of Eq.~\ref{tree-vev} and
reducing the degree-of-freedom count to only the NGB's. The
nonlinear constraints for $S$ are
    \beqs
    \Tr[SS^*]&=&s^2F^2 \label{NC1}\\
    \Tr[(S\times S)(S\times S)^*]&=&0
    \label{NC2}.
    \eeqs
With $S$ written in the form
   \beq
    S=\left(%
\begin{array}{ccc}
    s_{11} & s_{12} & s_{13} \\
    s_{12} & s_{22} & s_{23} \\
    s_{13} & s_{23} & s_{33} \\
\end{array}%
\right),
   \eeq
where the the $s_{ij}$ are complex fields,  Eq.~\ref{NC1} gives one
constraint for these 12 real fields.

Eq.~\ref{NC2} can be written in the form
   \beqs
&&\Tr[(S\times S)(S\times S)^*] \nonumber \\
   &=&|s_{11}s_{22}-s^2_{12}|^2+
|s_{11}s_{33}-s^2_{13}|^2+ |s_{22}s_{33}-s^2_{23}|^2 \nonumber \\
&&+2|s_{12}s_{13}-s_{11}s_{23}|^2+2|s_{12}s_{23}-s_{22}s_{13}|^2
\nonumber \\
&&+2|s_{13}s_{23}-s_{33}s_{12}|^2
\nonumber \\
&=&0. \eeqs Each of the absolute values must vanish, leading to a set of
three, independent complex equations ($6$ constraints in all) . They can be
taken to be \beqs
&& s_{11}s_{22}=s^2_{12} \\
&& s_{11}s_{33}=s^2_{13}\\
&& s_{22}s_{33}=s^2_{23}
   \eeqs
The same constraints apply to the elements of $\Sigma$.

The nonlinear constraint coupling $S$ and $\Sigma$ is:
   \beqs
   && \Tr[S\Sigma^*S^*\Sigma] \nonumber = 0\,. \eeqs
It can be written in the form \beqs &&
|s_{12}\sigma_{12}s_{23}\sigma_{23}+s_{12}\sigma_{12}s_{13}\sigma_{13}+s_{23}\sigma_{23}s_{13}\sigma_{13}|^2
   \nonumber \\
   &&\times
|s_{12}\sigma^*_{12}s_{23}\sigma^*_{23}+s_{12}\sigma^*_{12}s_{13}\sigma^*_{13}+s_{23}\sigma^*_{23}s_{13}\sigma^*_{13}|^2
   \nonumber \\
   &&/|s_{12}s_{13}s_{23}\sigma_{12}\sigma_{13}\sigma_{23}|^2 \nonumber \\
   &=&0\,, \eeqs
where $\sigma_{ij}$ are the elements of $\Sigma$. This form, which is a set
of $2$ constraints, is written making use of the separate nonlinear
constraints on $S$ and $\Sigma$. Thus the total number of constraints is
$16$, reducing the $24$ degrees of freedom in $S$ and $\Sigma$ to the $8$
NGB's of the visible sector.

These constraints also lead to the VEV's of Eq.~\ref{tree-vev}. If
we rotate $\langle S \rangle$ into diagonal form, then the
nonlinear constraints on $S$ allow only one non-vanishing element
which we take to be the $33$ element. The constraint coupling $S$
and $\Sigma$ then gives $\langle \sigma_{33}\rangle = 0$. But then
the above constraint equations on $S$, with $s_{ij}$ replaced by
$\sigma_{ij}$, demand that $\langle\sigma_{13}\rangle = \langle
\sigma_{23}\rangle  = 0$. Then $\langle \Sigma \rangle$ can be put
into diagonal form by an $SU(2)$ transformation leaving $\langle S
\rangle$ untouched. The constraints on $\Sigma$ demand that only
one element be nonzero, which we take to be the $22$ element.

We note that each of the above constraints can be derived from an
appropriate potential providing a phenomenological description of
the underlying dynamics~\cite{LFLi}. We assume here that they
emerge from the true underlying theory, the UV completion of our
EFT.

In a similar manner, the set of $3$ nonlinear constraints on the
hidden-sector $H$ is
    \beqs
   \Tr[HH^*]&=&O(1)\,F^
   2 \nonumber \\
   \Tr[(H\times H)(H\times H)^*]&=&O(1)\,F^4 \nonumber \\
   |{\rm det}\, H| &=& O(b^2)\,F^3,
    \eeqs
reducing the $12$ degrees of freedom in $H$ to the $9$ NGB's of
the hidden sector. There are now a total of $17$ NGB's Once the
visible and hidden sectors are linked by the family gauge
interaction and the third operator of ${\cal L}_Y^{\prime}$, all
but $9$ of these become PNGB's.

%%%%%%%%%%%%%%%%%%%%%%%%%%%%%%%%%%%%%%%%%%%%%%%%%%%%%%%%%%%%%%%%%%%%%%
%% Acknowledgments %%%%%%%%%%%%%%%%%%%%%%%%%%%%%%%%%%%%%%%%%%%%%%%%%%%
%%%%%%%%%%%%%%%%%%%%%%%%%%%%%%%%%%%%%%%%%%%%%%%%%%%%%%%%%%%%%%%%%%%%%%
\vspace{1.0cm}
\begin{acknowledgments}
This work was partially supported by Department of Energy grants
DE-FG02-92ER-40704 (T.A. and Y.B.) and DE-FG02-96ER40956 (M.P.). We thank
Michele Frigerio, Walter Goldberger, Adam Martin, Robert Shrock, and Witold
Skiba for useful discussions.
\end{acknowledgments}

%%%%%%%%%%%%%%%%%%%%%%%%%%%%%%%%%%%%%%%%%%%%%%%%%%%%%%%%%%%%%%%%%%%%%%
%%%  Bibliography  %%%%%%%%%%%%%%%%%%%%%%%%%%%%%%%%%%%%%%%%%%%%%%%%%%%
%%%%%%%%%%%%%%%%%%%%%%%%%%%%%%%%%%%%%%%%%%%%%%%%%%%%%%%%%%%%%%%%%%%%%%

\end{document}